\documentclass[pre,aps,twocolumn]{revtex4}

\usepackage{graphicx}
\usepackage{amsmath}
\usepackage{amssymb}
\usepackage{ifthen}
\usepackage{booktabs}
\usepackage{color}

\usepackage{graphicx}
\usepackage{dcolumn}
\usepackage{bm}
\usepackage{longtable}
\usepackage{gensymb}

\newcommand{\bb}{\begin{equation}}
\newcommand{\ee}{\end{equation}}
\newcommand{\ba}{\begin{eqnarray*}}
\newcommand{\ea}{\end{eqnarray*}}
\newcommand{\rhor}{\rho({\bf r})}
\newcommand{\dd}{{\rm d}}
\newcommand{\rr}{{\mathbf r}}
\newcommand{\dr}{{\rm d}{\bf r}}

\begin{document}

\title{Geometry-induced interface pinning at completely wet walls}

\author{Alexandr \surname{Malijevsk\'y}}
\affiliation{
{Department of Physical Chemistry, University of Chemical Technology Prague, Praha 6, 166 28, Czech Republic;}\\
 {Department of Molecular and Mesoscopic Modelling, ICPF of the Czech Academy Sciences, Prague, Czech Republic}}                

\begin{abstract}
\noindent We study complete wetting of solid walls that are patterned by parallel nanogrooves of depth $D$ and width $L$ with a periodicity of $2L$.
The wall is formed of a material which interacts with the fluid via a long-range potential and exhibits first-order wetting transition at temperature
$T_w$, should the wall is planar. Using a non-local density functional theory we show that at a fixed temperature $T>T_w$ the process of complete
wetting depends sensitively on two microscopic length-scales $L_c^+$ and $L_c^-$. If the corrugation parameter $L$ is greater than $L_c^+$, the
process is continuous similar to complete wetting on a planar wall.  For $L_c^-<L<L_c^+$, the complete wetting exhibits first-order \emph{depinning
transition} corresponding to an abrupt unbinding of the liquid-gas interface from the wall.  Finally, for $L<L_c^-$ the interface remains pinned at
the wall even at bulk liquid-gas coexistence. This implies that nano-modification of substrate surfaces can always change their wetting character
from hydrophilic into hydrophobic, in direct contrast to the macroscopic Wenzel law. The resulting surface phase diagram reveals close analogy
between the depinning and prewetting transitions including the nature of their critical points.
\end{abstract}

\maketitle


The recent advances in nanophysics have not only revealed promising possibilities in modern technologies but also induced new theoretical challenges.
This includes a particularly important problem that has attracted enormous interest across different scientific branches and which can be formulated
as follows: what is the effect of a solid surface structure on its wetting properties? On a macroscopic level, the influence of a non-planar
structure on adsorption behaviour of the substrate can be described by  Wenzel's law \cite{wenzel36}
 \bb
  \cos\theta^*=r\cos\theta\,, \label{wenzel}
 \ee
which relates Young's contact angle $\theta$ of a liquid droplet on a planar surface with an apparent contact angle $\theta^*$ of a liquid droplet on
a structured surface. Since the roughness parameter $r>1$, Eq.~(\ref{wenzel}) implies that $\theta^*>\theta$ if $\theta>\pi/2$ and $\theta^*<\theta$
if $\theta<\pi/2$ meaning that the wetting/drying properties are amplified by the surface structure.

\begin{figure}
\includegraphics[width=8cm]{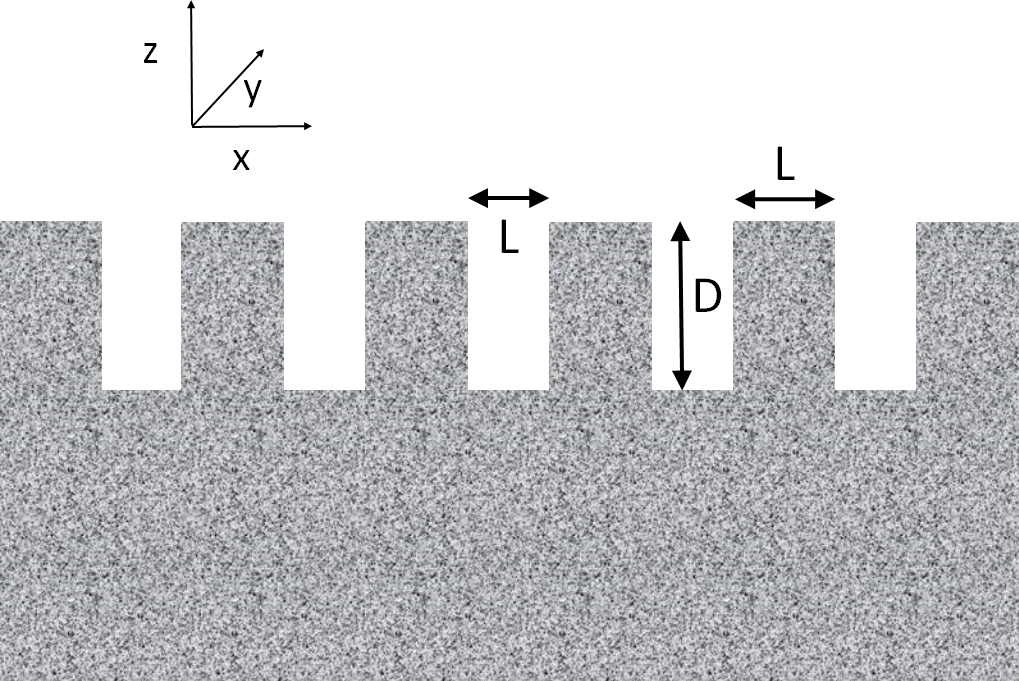}
\caption{Sketch of a cross-section  of the model substrate with microscopic grooves of depth $D$ and width $L$.  The system is assumed to be periodic
along the $x$-axis with a periodicity of $P=2L$ and translation invariant along the $y$-axis. } \label{sketch}
\end{figure}

Nevertheless, it turns out that the behaviour of fluids adsorbed at rough substrates is, at a minimum, much more complex \cite{netz, chow, swain,
quere1, quere2, herminghaus_rev, erbil, krasimir}, especially upon decreasing the length-scale characterizing structure of the solid surface
\cite{fox, ruckenstein, rauscher, gross, herminghaus, dutka, malanoski}. Further aspects, such that the surface geometry, the nature of microscopic
forces, line tension contribution, strong packing effects that fluid molecules experience near the wall etc., are all ignored in the phenomenological
Wenzel law, yet they are crucial for obtaining the full picture. In particular, it has been recognized that interfacial phenomena occurring on
structured substrates depend on the nature of wetting properties of the pertinent planar substrate \cite{santori, rascon, rejmer0, kubalski,
rejmer02, rejmer07, silvestre, rodriguez}: If the planar substrate experiences continuous (critical) wetting transition  \cite{dietrich, schick,
bonn} at temperature $T_w$ (and at bulk liquid-gas coexistence) which corresponds to vanishing of the contact angle, $\theta(T_w)=0$, the corrugated
substrate also exhibits continuous wetting transition at the same temperature. The character of the wetting transition is also unchanged if the
planar substrate exhibits first-order wetting but in this case the wetting temperature of the corrugated substrate is shifted towards lower values,
qualitatively in line with Eq.~(\ref{wenzel}) \cite{rascon, kubalski, rejmer02}. Moreover, the wetting can be preceded by unbending (or filling)
transition corresponding to an abrupt condensation of the fluid inside the wall troughs, provided the corrugation amplitude is sufficiently large.

\begin{figure}
\includegraphics[width=8cm]{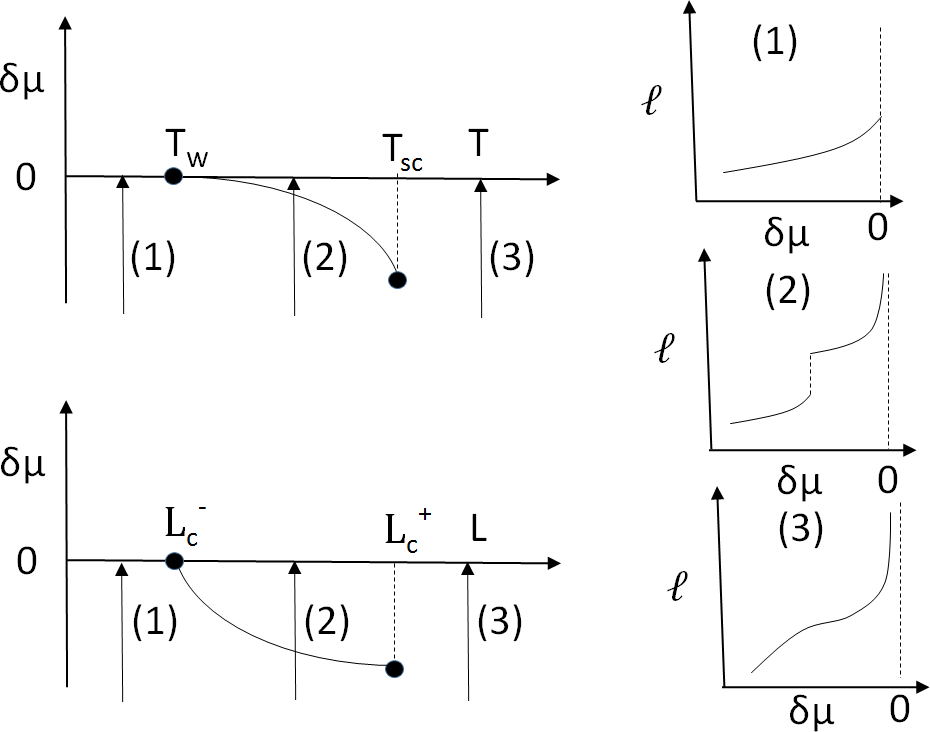}
\caption{A comparison between surface phase diagrams for a planar wall exhibiting first-order wetting transition at temperature $T_w$ (top left) and
a periodically corrugated wall at a fixed temperature $T>T_w$ (bottom left). Also shown are three adsorption isotherms corresponding to thermodynamic
paths denoted in the phase diagrams.} \label{pd_scheme}
\end{figure}


These predictions are based on a mesoscopic analysis of the liquid-gas interface interacting with the solid wall according to an effective (binding)
potential \cite{dietrich}. However, if the substrate structure is microscopic, i.e. on the scale of molecular diameters, a more detailed treatment of
an adsorbed fluid is needed. In this work, we consider a model substrate formed by a solid planar wall into which a one-dimensional array of
rectangular grooves of depth $D$ and width $L$ is etched with a periodicity $P=2L$. The groove parameters $D$ and $L$ are deemed to be
microscopically small, while the length of the grooves $L_y$ along the remaining Cartesian axis is assumed to be macroscopically long, so that the
wall corrugation breaks the translation symmetry of the system only in one direction (cf. Fig.~\ref{sketch}). The wall is at contact with a bulk gas
at a (subcritical) temperature $T>T_w$ and the chemical potential $\mu<\mu_{\rm sat}(T)$, where $T_w$ is the wetting temperature at which first-order
wetting transition occurs at the corresponding planar wall and $\mu_{\rm sat}(T)$ is the chemical potential at bulk liquid-gas coexistence. For
planar walls the process $\mu\to\mu_{\rm sat}(T)$ is known as complete wetting and can be characterized by a an unbinding of the liquid-gas interface
mean height $\ell$ which eventually diverges according to the power-law $\ell\sim|\delta\mu|^{-\beta_{co}}$ where $\delta\mu=\mu-\mu_{\rm sat}$ and
the critical exponent $\beta_{co}=1/3$ for systems with long-ranged (dispersion) forces \cite{dietrich, schick, bonn}. We also recall that for
first-order wetting the singular behaviour of the surface free energy at $T_w$ extends to $T>T_w$ and below $\mu_{\rm sat}$, giving rise to a finite
jump in $\ell(\mu)$; this prewetting transition terminates at the surface critical point $T_{sc}$. Schematically, the surface phase diagram for
complete wetting at a planar wall is displayed in Fig.~\ref{pd_scheme}, together with three illustrative adsorption isotherms. Here we also show the
$L$-$\delta\mu$ phase diagram corresponding to our model substrate which summarizes the main results of this work. It demonstrates three possible
adsorption scenarios, depending on $L$: i) the adsorption is continuous similar to complete wetting on a planar wall if the corrugation parameter $L$
is greater than a certain critical value $L_c^+$; ii) for $L$ below $L_c^+$ but above $L_c^-$ the adsorption exhibits first-order \emph{depinning
transition} at some value $\mu(L)<\mu_{\rm sat}$ corresponding to an abrupt depinning of the interface from the wall followed by a continuous
divergence of the interface height as $\mu\to\mu_{\rm sat}$; iii) finally, for $L<L_c^-$ the interface remains pinned at the wall even at $\mu_{\rm
sat}$ preventing complete wetting despite Young's contact angle $\theta=0$.

We have obtained our results using a microscopic density functional theory (DFT) \cite{evans79}, which has proven to be an extraordinary useful tool
for a description of structure and phase behaviour of inhomogeneous fluids \cite{evans16}. All the information about the given model fluid properties
is embraced in the intrinsic free energy functional $F[\rho]$ of the local one-body fluid density $\rhor$ which, for simple fluids, can be decomposed
in a perturbative manner as follows:
 \bb
 F[\rho]=F_{\rm id}[\rho]+F_{\rm rep}[\rho]+F_{\rm att}[\rho]\,.
 \ee
Here, $F_{\rm id}=\beta^{-1}\int\dr\rho(\rr)\left[\ln(\rhor\Lambda^3)-1\right]$ is the kinetic (ideal gas) contribution to the free energy where
$\beta=1/k_BT$ is the inverse temperature and $\Lambda$ is the thermal de Broglie wavelength. The excess part of the free energy functional due to
the fluid-fluid interactions is further separated to the repulsive portion $F_{\rm rep}$, which is mapped onto a system of hard spheres with a
diameter $\sigma$  within the non-local Rosenfeld fundamental-measure-theory functional \cite{ros}, and the attractive part which we treat in the
mean-field manner $F_{\rm att}= \frac{1}{2}\int\int\dd\rr\dd\rr'\rhor\rho(\rr')u_{\rm att}(|\rr-\rr'|)$. For the attractive part of the fluid-fluid
interaction, $u_{\rm att}(r)$, we have chosen the truncated and non-shifted Lennard-Jones-like potential
 \bb
 u_{\rm att}(r)=\left\{\begin{array}{cc}
 0\,;&r<\sigma\,,\\
-4\varepsilon\left(\frac{\sigma}{r}\right)^6\,;& \sigma<r<r_c\,,\\
0\,;&r>r_c\,,
\end{array}\right.\label{ua}
 \ee
where the parameters $\varepsilon$ and $\sigma$ are used as the energy and length-scale units, respectively, and where the potential cut-off is set
to $r_c=2.5\,\sigma$. The microscopic model accounts accurately for the short-ranged fluid correlations (and thus the packing effects) and satisfies
exact statistical mechanical sum rules \cite{hend92}. The confining wall (illustrated in Fig.~1), steps into the theory within the external potential
$V(\rr)=V(x,z)$ which is obtained by integrating the wall-fluid atom-atom interactions $\phi_w(r)$ over the whole volume of the wall. With the wall
atoms assumed to be distributed uniformly with a density $\rho_w$ and interacting with the fluid atoms via the Lennard-Jones (LJ) potential
$\phi_w(r)=4\varepsilon_w\left[\left(\sigma/r\right)^{12}-\left(\sigma/r\right)^{6}\right]$, the wall potential can be expressed as
 \bb
 V(\rr)=V_\pi(z)+\sum_{n=-\infty}^\infty V_D(x+2nL,z)\,, \label{wall_pot}
 \ee
except for the region  corresponding to the domain of the wall in which case $V(x,z)=\infty$ as the wall is impenetrable. The potential $V_D(x,z)$ of
a single pillar of height $D$ and width $L$ can be split into the repulsive and attractive contributions $V_D(x,z)=V_6(x,z)+V_{12}(x,z)$ where
 \begin{eqnarray}
V_6(x,z)&=&-\frac{\pi}{3}\varepsilon_w\sigma^6\rho_w\left[\psi_6(x,z)-\psi_6(x,z-D)\right.\nonumber\\
&&\left.-\psi_6(x-L,z)+\psi_6(x-L,z-D)\right]
 \end{eqnarray}
 and
 \begin{eqnarray}
V_{12}(x,z)&=&\pi\varepsilon_w\sigma^{12}\rho_w\left[\psi_{12}(x,z)-\psi_{12}(x,z-D)\right.\nonumber\\
&&\left.-\psi_{12}(x-L,z)+\psi_{12}(x-L,z-D)\right]
 \end{eqnarray}
 with
  \bb
\psi_6(x,z)=\frac{2x^4+x^2z^2+2z^4}{2x^3z^3\sqrt{x^2+z^2}} 
 \ee
 and
  \begin{widetext}
 \bb
\psi_{12}(x,z)=\frac{1}{128}{\frac {128\,{x}^{16}+448\,{x}^{14}{z}^{2}+560\,{x}^{12}{z}^{4}+280\,{
x}^{10}{z}^{6}+35\,{x}^{8}{z}^{8}+280\,{x}^{6}{z}^{10}+560\,{x}^{4}{z} ^{12}+448\,{z}^{14}{x}^{2}+128\,{z}^{16}}{{z}^{9}{x}^{9} \left( {x}^{2
}+{z}^{2} \right) ^{7/2}}} -\frac{1}{z^9}\,.
 \ee
 \end{widetext}
 Finally, the potential $V_\pi(z)$ in (\ref{wall_pot}) is the standard $9$-$3$ LJ potential induced by a planar wall spanning the volume $z<0$.

 \begin{figure}
 \centerline{\includegraphics[width=8cm]{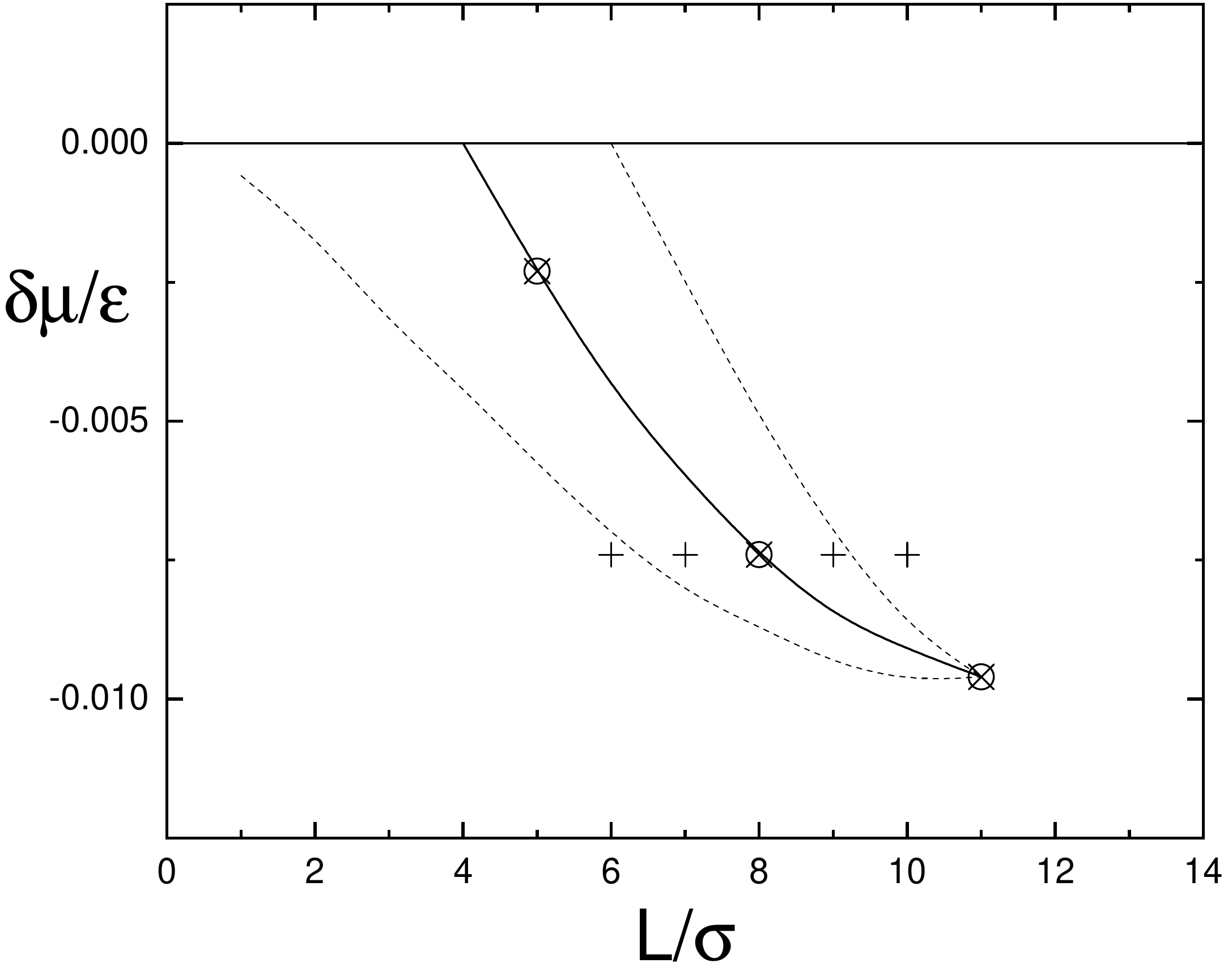}}
 \caption{The surface phase diagram as obtained from DFT for $T=0.92\,T_c$ and $D=10\,\sigma$. The line displays the loci of the first-order depinning
transition and terminates at the critical distance $L_c^+\approx11\,\sigma$ above which the transition is continuous. Below $L_c^-\approx4\,\sigma$
the gas-liquid interface remains pinned at the wall (which is thus not wet) even at saturation $\delta\mu=0$ represented by the horizontal line. Also
shown are the spinodal lines (dashed) which denote the limit of stability of the depinned state (the lower one) and the pinned state (the upper one).
The symbols $\times$  ($+$) indicate the points for which the corresponding binding potentials are plotted in Fig.~5 (Fig.~6). \label{pd}}
\end{figure}

\begin{figure}
\centerline{\includegraphics[width=4.5cm]{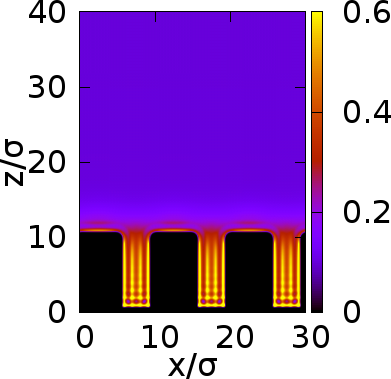}\hspace*{0.2cm}\includegraphics[width=4.5cm]{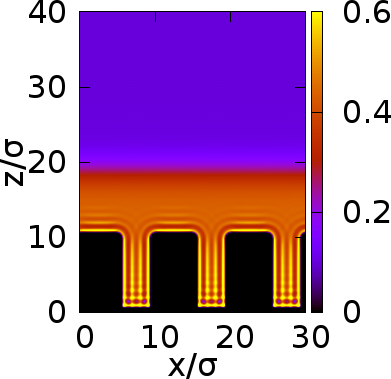}}

\vspace*{0.5cm}

\centerline{\includegraphics[width=4.5cm]{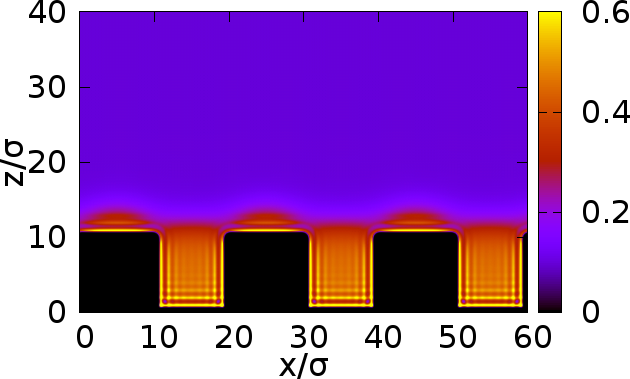}\hspace*{0.2cm}\includegraphics[width=4.5cm]{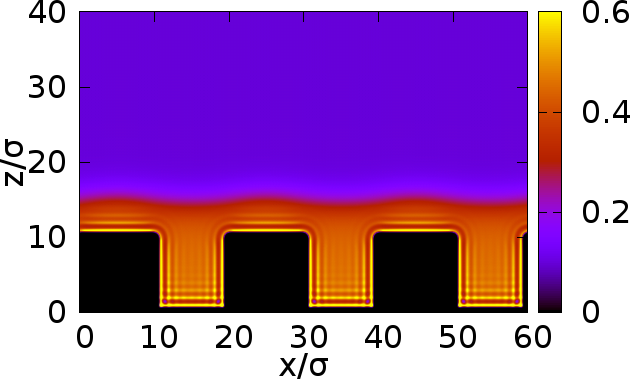}}
 \caption{Equilibrium density profiles of pinned (left)
and depinned (right) coexisting wetting states for $L=5\,\sigma$ (upper panels) and $L=10\,\sigma$ (bottom panels).} \label{dp}
\end{figure}

The equilibrium density profile is obtained by minimizing the grand potential functional
 \bb
 \Omega[\rhor]=F[\rhor]+\int\rhor (V(\rr)-\mu)\dr \label{omega}
 \ee
which is solved iteratively on a two-dimensional grid with a uniform spacing of $0.1\,\sigma$. The minimization determines the equilibrium density
profile $\rhor$ and also the thermodynamic grand potential $\Omega={\rm min}\,\Omega[\rhor]$. Having set the wall parameter
$\varepsilon_w\rho_w=\varepsilon/\sigma^3$ we find the wetting temperature of the planar wall $T_w\approx0.8\,T_c$, where $T_c$ is the bulk critical
temperature of the fluid.

In Fig.~\ref{pd} we display the surface phase diagram in the $L$-$\delta\mu$ projection obtained from the DFT for $T=0.92\,T_c>T_w$ and the grooves
depth $D=10\,\sigma$. The phase diagram shows the line corresponding to first-order depinning transition at which the grand potentials obtained by
minimizing of (\ref{omega}) for the pinned (low adsorption) and depinned (high adsorption) states are equal. The line connects the (horizontal) bulk
coexistence line at $L_c^-\approx4\,\sigma$ meaning that below this threshold the wall structure prevents complete wetting. The depinning line
terminates at the critical point $L_c^+\approx11\,\sigma$ above which the depinning is continuous. For $L>L_c^+$, a unique solution for the density
profile is always obtained regardless of the initial state which the minimization of (\ref{omega}) starts from. However, below $L_c$, the function
$\Omega(L)$ exhibits two local minima within the interval constrained by the spinodal lines which are also shown in Fig.~\ref{pd}. Therefore, this
interval indicates a range of metastable extensions of either solution characteristic to first-order transitions. The spinodals display the loci
where one of the two minima in $\Omega(L)$ vanishes and becomes an inflection (cf. Fig.~\ref{bp_974} below), i.e. the limit of stability of the
corresponding (pinned or depinned) configuration.

To illustrate the change in the structure of the fluid at the depinning transition we show in Fig.~\ref{dp} density profiles (over three periods)
corresponding to coexisting pinned and depinned states for $L=5\,\sigma$ and $L=10\,\sigma$. The two examples differ rather remarkably; for low $L$,
the depinned state possesses a thick wetting layer with essentially flat interface, while the upper part of the wall is only microscopically wet in
the pinned state. For large $L$, the width of the wetting film after the transition is substantially smaller and exhibits distinct periodic
corrugation of the liquid-gas interface which follows closely the lateral inhomogeneity in the wall potential. Before the transition, liquid droplets
that are pinned at the wall edges are now present, as the width of the pillars is large enough to accommodate them. One should also notice the
strongly inhomogeneous fluid structure in the grooves showing pronounced layering that get connected at the depinned states; this suggests that the
transition can also be viewed as bridging of the condensed phase filling the grooves.

\begin{figure}
\centerline{\includegraphics[width=9cm]{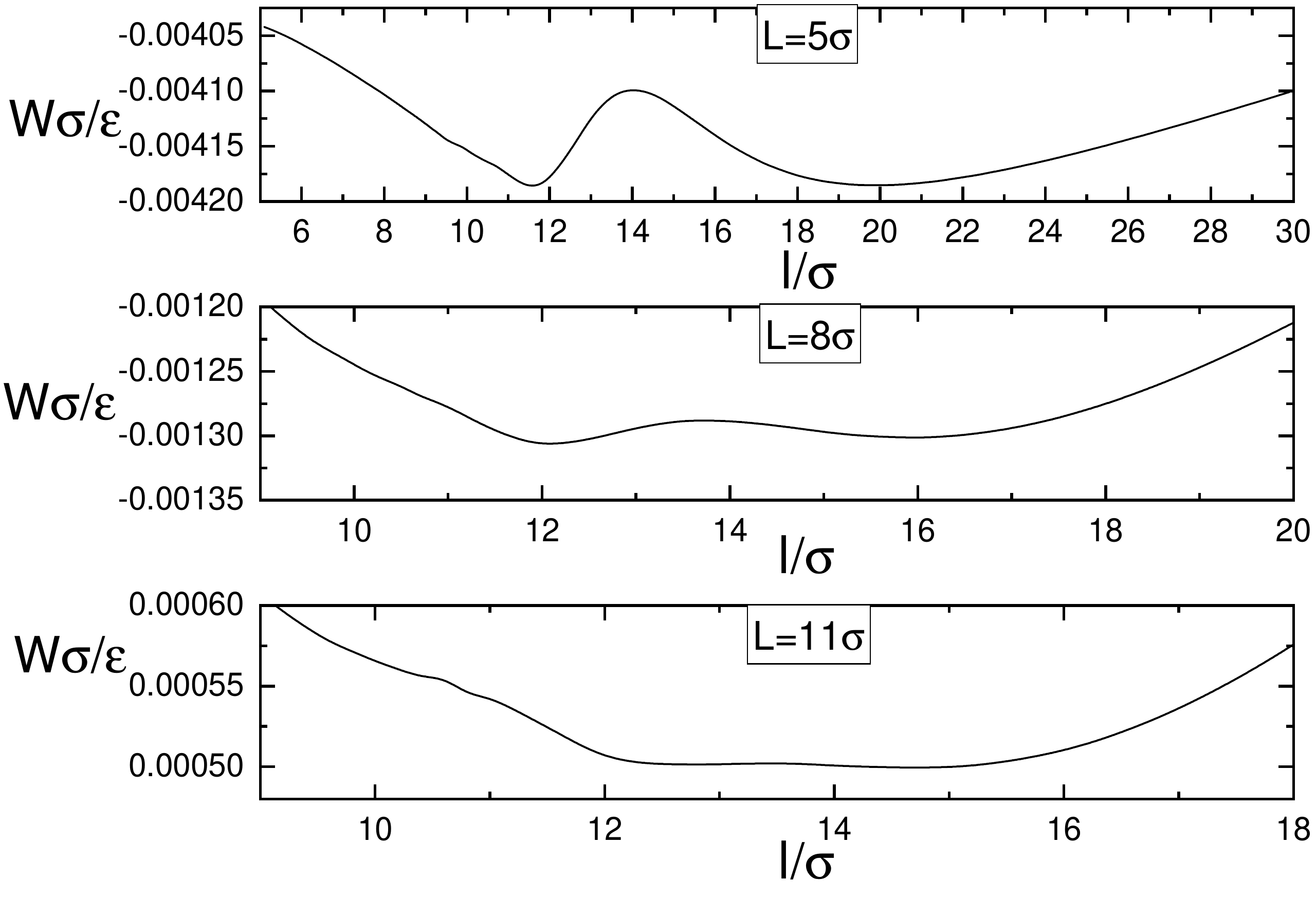}}
 \caption{Binding potentials for (a) $L=5\,\sigma$, (b) $L=8\,\sigma$ and (c) $L=11\,\sigma$,
 corresponding to the coexistence of the pinned and the depinned state. In all the cases, the binding potential exhibits two local minima
 of the same depth
 which essentially merge when $L=11\,\sigma$ indicating a close proximity to the critical point $L_c^+$.}
  \label{bp_eq}
\end{figure}

\begin{figure}
 \centerline{\includegraphics[width=9cm]{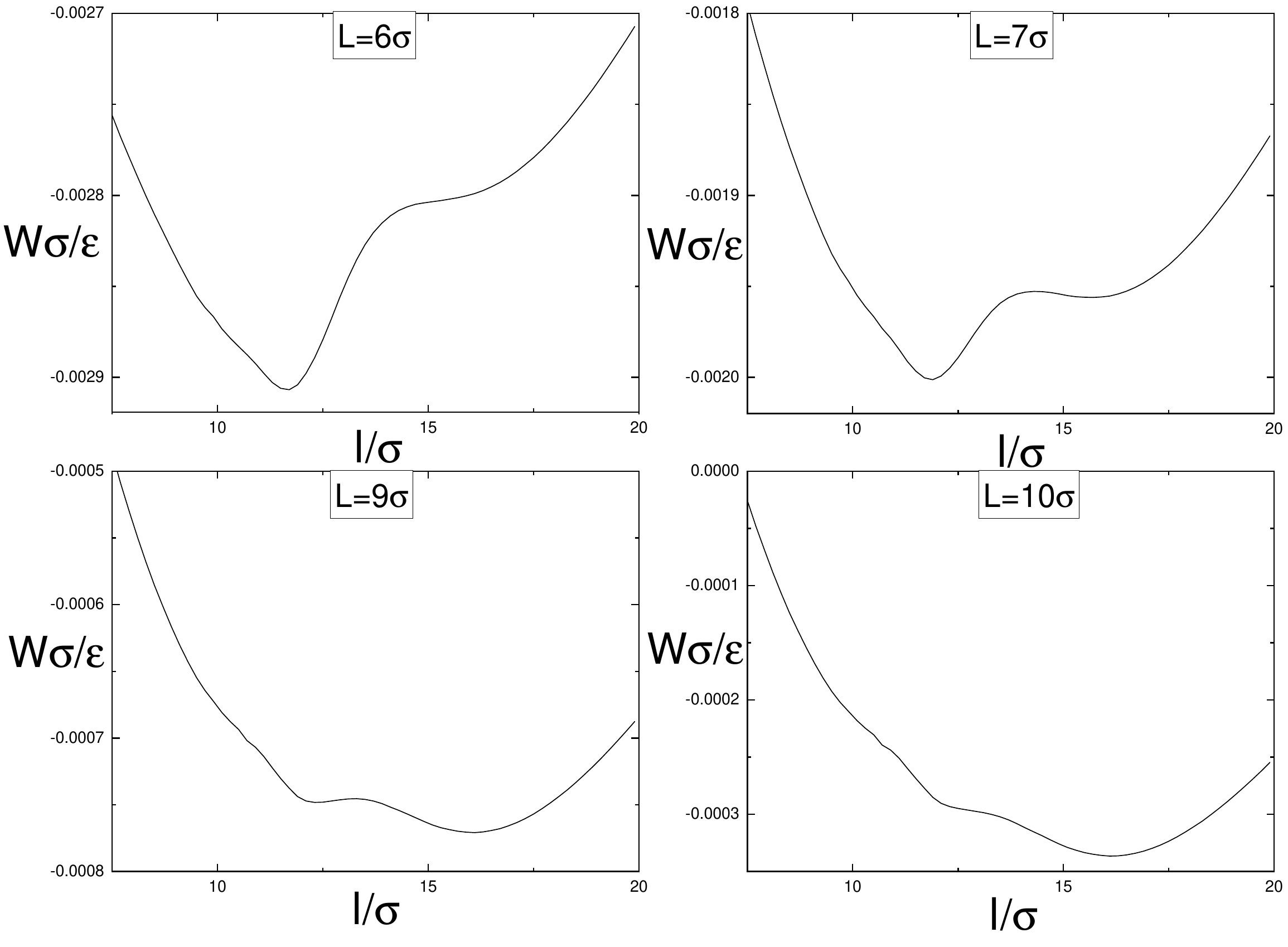}}
\caption{Binding potentials for various values of $L$: (a) $L=6\,\sigma$, (b) $L=7\,\sigma$, (c) $L=9\,\sigma$ and (d) $L=10\,\sigma$,  as obtained
from DFT at the fixed chemical potential $\mu=-3.98\,\varepsilon$ corresponding to the depinning transition for $L=8\,\sigma$. For the lowest
($L=6\,\sigma$)  and the highest ($L=10\,\sigma$) value of the corrugation parameter displayed, one of the two local minima disappears suggesting
that these states are already beyond the spinodal points of the transition (cf. Fig.~\ref{pd}).} \label{bp_974}
\end{figure}

Some more details about the depinning transition can be obtained by constructing the binding potential $W(\ell)$, i.e. the constrained free energy
per unit length at the fixed mean height of the adsorbed film $\ell$ (from the grooves bottom), $W(\ell)={\rm
min}_{\rhor}\left.\Omega[\rhor]\right|_{\ell}/L_y$, so that the mean-field equilibrium state is given by the global minimum of $W(\ell)$. For this,
we minimize the grand potential (\ref{omega}) as a subject of fixed adsorption \cite{bind} $\Gamma=\int\dd x\int\dd z (\rho(x,z)-\rho_b)$ where
$\rho_b(\mu)$ is the bulk fluid density. In Fig.~\ref{bp_eq} we display the binding potentials corresponding to the three points laying on the
depinning line as depicted in Fig.~\ref{pd}. For the lowest value of $L$, the binding potential possesses two distinct local minima of the same
depths that are separated by a well pronounced free-energy barrier. On increasing $L$, i.e. by approaching $L_c^+$, the free-energy barrier is
lowered, as well as the gap between the two minima which eventually merge at the critical point. Clearly, this is only the second minimum of
$W(\ell)$ the position of which depends on $L$, with the first minimum being always located near $D$ corresponding to the grooves top.


\begin{figure}
\centerline{\includegraphics[width=8cm]{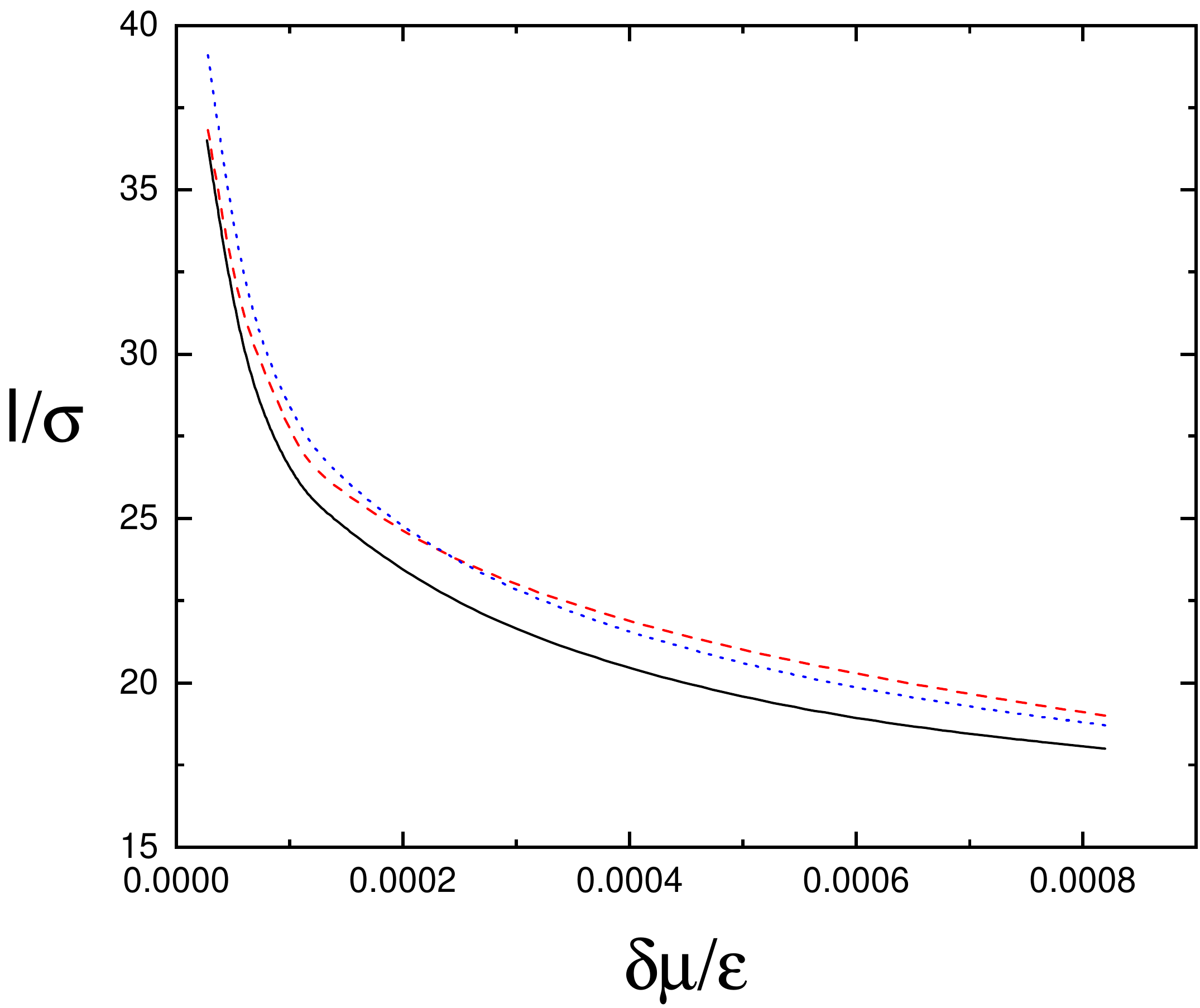}}
 \caption{DFT results for the
 height of the wetting layers at the structured wall (black solid line) with $D=10\,\sigma$ and $L=5\,\sigma$, and the planar wall (red dashed line)
 with the potential strength as given by Eq.~(\ref{eq_cov}). Also shown is the result obtained by a direct solution of
 Eq.~(\ref{eq_cov}) (blue dotted line).}
 \label{cov}
\end{figure}

In Fig.~\ref{bp_974} we also show the binding potentials for the points away of the depinning transition at the fixed chemical potential. The effect
of varying $L$ is now different; namely it determines the depths of the two competing free-energy minima but their locations remain unchanged within
the whole interval of $L$ between the spinodal points (cf. Fig.~\ref{pd}) beyond which only one local minimum in $W(\ell)$ exists. This can be
explained as follows. Within the sharp-kink approximation \cite{dietrich}, the binding potential of the system in the depinned state can be written
as
 \bb
 W(\ell)=2(\ell-D)L\delta\mu\Delta\rho+\frac{AL}{(\ell-D)^2}+\frac{AL}{\ell^2}\;\;\;(\ell>D)\,. \label{bp_groove}
 \ee
Here, the first term on the right hand side is the free-energy cost for the presence of the metastable liquid (ignoring the constant contribution due
to the filled grooves) and the remaining terms account for the local effective interaction between the wall and the liquid-gas interface which is
considered to be flat. This interaction is approximated by a simple quadratic power-law in the inverse height of the interface with the amplitude
(Hamaker constant) $A=\pi\Delta\rho\rho_w\varepsilon_w\sigma^6/3$, where $\Delta\rho$ is the difference between the bulk coexisting densities.
Therefore, within the approximation $W(\ell)$ scales linearly with $L$ and hence the interface height is $L$-independent, in line with the DFT
results. It should be emphasized that the model (\ref{bp_groove}) is only applicable to the depinned state (and is meaningful only above the lower
spinodal of the phase diagram) and has no relevance to the nature of the depinning transition. Indeed, the full model would rely on a definition of
two distinct fluid configurations (pinned and depinned states) and would not thus be able to predict the presence of the critical point $L_c^+$.
However, Eq.~(\ref{bp_groove}) can be further used to find a correspondence between our substrate model involving grooves and a simple planar wall
covered by a wetting film of width $\ell_\pi$, for which the dominating contribution to the binding potential is
 \bb
 W_\pi(\ell_\pi)=\delta\mu\Delta\rho\ell_\pi +\frac{A'}{\ell_\pi^2}\,, \label{bp_plane}
 \ee
defined per unit area. The effective Hamaker constant $A'$ of the planar wall is determined by equating the heights of the wetting films at the
corresponding substrates, $\ell(\delta\mu)=\ell_\pi(\delta\mu)$, obtained by minimization of (\ref{bp_groove}) and (\ref{bp_plane}), respectively.
This leads to the condition
 \bb
 \left(1-\frac{1}{\tilde{\ell}}\right)^3=\frac{y}{2\tilde{\ell}^3-y}\,, \label{eq_cov}
 \ee
for $\tilde{\ell}=\ell/D$ as a function of the scaling parameter $y=2A/(\delta\mu\Delta\rho D^3)$, which eventually yields $A'=\tilde{\ell}^3A/y$. It
can be checked easily, that $A'\to\infty$ for $D\to\infty$ and $A'\to A$ for $D\to0$, as expected \cite{covar}. Using DFT, we test this result by
comparing the height of the depinned interface at the structured wall with $D=10\,\sigma$ and $L=5\,\sigma$ and the potential strength
$\varepsilon_w=\varepsilon$, with that corresponding to the planar wall with the potential strength $\varepsilon'_w/\varepsilon_w=A'/A$ as obtained
from Eq.~(\ref{eq_cov}). The comparison shown in Fig.~\ref{cov} reveals that the two solutions, $\ell$ and $\ell_\pi$, are fairly close to each
other, although the interface height above the planar wall $\ell_\pi$ is systematically slightly larger. For completeness, we also plot
$\ell(\delta\mu)$ as obtained directly from Eq.~(\ref{eq_cov}) which almost follows $\ell_\pi$; the upshot is that while the asymptotic form of the
planar binding potential (\ref{bp_plane}) works very accurately within the displayed interval, the binding potential for the structured wall
(\ref{bp_groove}) provides still a reasonable approximation.

In summary, we have studied adsorption of a solid wall structured by a linear array of parallel rectangular grooves above the wetting temperature, so
that Young's contact angle of the wall is zero. We have found that the presence of the wall structure does not qualitatively change the process of
complete wetting unless the characteristic length of the wall structure $L$ is microscopically small.
In this case, the mechanism of complete wetting is via bridging of liquid layers inside the grooves over the top of the wall. The corresponding free
energy change involves a contribution associated with the line tension $\tau$ pertinent to a contact of the liquid-gas interface with the wall edges.
This term competes within the excess free energy with the surface tension effects as $\tau/L$ and is thus increasingly relevant as $L$ decreases. Its
role becomes dominant for $L<L_c^-$ such that the free energy barrier cannot be overcome even at the bulk coexistence $\delta\mu=0$ and the bridging
and complete wetting of the wall is thus not possible. However, as our DFT calculations show, this is only in the case when $L$ is less than about
four molecular diameters ($\sigma$). For larger $L$, the system experiences competition between two free energy minima giving rise to first-order
depinning transition. The edge effects and hence the transition is gradually weaker as $L$ increases and eventually becomes continuous at
$L=L_c^-(\approx11\,\sigma$) where the barrier vanishes. For $L\ge L_c^+$ the shape of the interface changes smoothly all the way along its unbinding
from the wall. The $\delta\mu$-$L$ phase diagram reveals some analogy between prewetting and depinning transitions, although the curvatures of the
corresponding lines have opposite signs. This study can be extended in numerous ways. A natural generalization of our model would treat the grooves
width $L$ and the periodicity $P$ as independent parameters; this, for example, would convert Eq.~(\ref{eq_cov}) simply into
$(1-1/\tilde{l}^3)=xy/(\tilde{\ell}^3+(x-1)y)$ with $x=L/P$ but the phase behaviour at such a wall can be expected to be considerably more
complicated. Further, here we have deliberately chosen a sufficiently high temperature in order to avoid prewetting at the wall; for $T<T_{sc}$ one
expects competition between depinning and prewetting. It would also be interesting to explore $D$-dependence of the depinning phenomena and check
possible scaling properties as in Eq.~(\ref{eq_cov}).
%
Finally, it should be noted that in view of its pseudo-2D character and the presence of long-range forces, the depinning transition would
not be rounded beyond the current mean-field analysis due to thermal fluctuations and should thus be accessible in real experiments.

\begin{acknowledgments}
\noindent This work was financially supported by the Czech Science Foundation, Project No. GA17-25100S.
\end{acknowledgments}


\begin{thebibliography}{99}
\bibitem{covar}
 If the correspondence is defined by a condition $\ell_\pi=\ell-D$ by considering the upper part of the structured wall as the
 reference level, the minus sign on the l.h.s. of Eq.~(\ref{eq_cov}) would change and this would also imply $A'\to A/2$ in the limit of
 $D\to\infty$.
\end{thebibliography}
\end{document}